\documentclass{webofc}
\usepackage[varg]{txfonts}   % Web of Conferences font
\usepackage{amsmath}
\usepackage{braket}
\usepackage{bm,color}
\begin{document}
\title{Structure of two- and three-alpha systems in cold neutron matter}

\author{\firstname{Hajime} \lastname{Moriya}\inst{1}\fnsep\thanks{\email{moriya@nucl.sci.hokudai.ac.jp}} \and
  \firstname{Hiroyuki} \lastname{Tajima}\inst{2,3}\fnsep
  %\thanks{\email{hiroyuki.tajima@phys.s.u-tokyo.ac.jp}}
  \and
  \firstname{Wataru} \lastname{Horiuchi}\inst{1}\fnsep
  %\thanks{\email{whoriuchi@nucl.sci.hokudai.ac.jp}}
  \and
  \firstname{Kei} \lastname{Iida}\inst{3}\fnsep
  %\thanks{\email{iida@kochi-u.ac.jp}}
  \and
  \firstname{Eiji} \lastname{Nakano}\inst{3}\fnsep
  %\thanks{\email{e.nakano@kochi-u.ac.jp}}
        % etc.
}

\institute{Department of Physics, Hokkaido University, Sapporo 060-0810, Japan
\and
           Department of Physics, Graduate School of Science, The University of Tokyo, Tokyo 113-0033, Japan
\and
           Department of Mathematics and Physics, Kochi Univerisity, Kochi 780-8520, Japan
          }

\abstract{
We present stability and structure 
%changes 
of two- and three-alpha systems embedded in 
dilute cold neutron matter.
By solving a few-alpha Schr\"{o}dinger equation
with quasiparticle properties, i.e., effective mass and 
induced two- and three-alpha interactions, which are evaluated 
in terms of Fermi polarons,
it is shown that $^8$Be and the Hoyle state become bound 
at densities of about
$10^{-4}$ and $10^{-3}$ of the saturation density, respectively.
It is also seen that, under cold neutron matter environment,
both systems become smaller
than the corresponding systems in vacuum.
Our results would affect astrophysical models for 
stellar collapse and neutron star mergers, 
as well as relevant reaction rates for nucleosynthesis.
}
\maketitle
\section{Introduction}
\label{introduction}
An alpha ($^4$He nucleus) cluster plays an important role 
in the nuclear structure of light nuclei
and astrophysical reactions relevant for the evolution of
chemical elements in a burning star.
In particular, the first $J^\pi=0^+$ excited state of $^{12}$C,
the so-called Hoyle state, is one of the most famous examples
in which three alpha clusters are well developed.
The Hoyle state works in such a way as to accelerate the 
triple-alpha reaction at low energies near the Gamow 
window~\cite{Hoyle54}, where the $^{12}$C element is generated 
through sequential reactions via the resonant ground state of $^8$Be. 
It is important to describe such alpha-induced reactions accurately
in the astrophysical explosive phenomena, e.g., core-collapse 
supernovae and neutron star mergers \cite{Oertel17}.

Recently, quasiparticle properties
of an alpha particle in dilute cold neutron matter
were discussed~\cite{Nakano20},
where the alpha particle in the neutron matter
was described as an impurity in a many-body background
like a Fermi polaron in ultracold atomic physics~\cite{Chevy10}.
In the polaron picture,
the effective mass of the alpha particle is controlled by
the interaction from the background.
In this paper we discuss possible changes in the structure
of two- and three-alpha systems
in dilute neutron matter at zero temperature.
By solving a few-alpha Schr\"{o}dinger equation precisely, we discuss the 
stability and structure changes of the ground state of $^8$Be
and the Hoyle state in such neutron matter.
In addition to the change of the effective mass,
we introduce two- and three-alpha interactions
induced by the neutron medium.
Since all the details of the present evaluation are described
in Ref.~\cite{Moriya21}, in this paper, we briefly recapitulate
those discussions and present some new results that were not presented there.

\section{Method}

We employ a multi-alpha cluster model to describe
the two- and three-alpha systems.
The effective Hamiltonian for the three-alpha system is given by
\begin{equation}
  H=\sum_{i} \frac{\bm{p}_{i}^2}{2M^*}-T_{cm} 
  + \sum_{i>j} \left(V_{\alpha \alpha,ij}+ V_{\mathrm{eff},ij}^{(2)}\right)
  + V_{\alpha \alpha \alpha} + V_{\mathrm{eff}}^{(3)}, 
\end{equation}
where $M^*$ is the effective mass of the alpha particle,
$\bm{p}_i$ is the momentum operator of
the $i$th alpha particle, $T_{cm}$ is the kinetic energy 
of the center-of-mass motion, $V_{\alpha \alpha}$ and $V_{\alpha \alpha \alpha}$
are two- and three-alpha interactions, respectively.
This two-alpha potential is deep enough to have some redundant 
or forbidden bound states
by the Pauli principle between alpha particles.
Details of the parameter set are given in Refs.~\cite{Fukatsu92,Kurokawa05}.
Physical states should be orthogonal to such forbidden states.
These states are eliminated practically
by using the projection method~\cite{Kukulin78}.

The effective mass of the alpha particle $M^*$
was given as a function of the Fermi momentum $k_F$ in Ref.~\cite{Nakano20}.
Here, medium induced two- and three-alpha interactions
($V_{\mathrm{eff}}^{(2)}$ and $V_{\mathrm{eff}}^{(3)}$) are also introduced
by considering the simple Feynman diagrams in which few-alpha clusters
exchange momenta. All the details are described in Ref.~\cite{Moriya21}.
The wave function of the few-alpha system is described by a superposition 
of fully symmetrized correlated Gaussian basis functions whose
variational parameters are optimized by 
the stochastic variational method \cite{Varga95,Suzuki98B}.
All the numerical setups are detailed in Refs.~\cite{Phyu20,Moriya21}.

\section{Results and discussion}

\begin{figure*}[h]
\includegraphics[bb=0 0 162 126, width=0.5\linewidth]{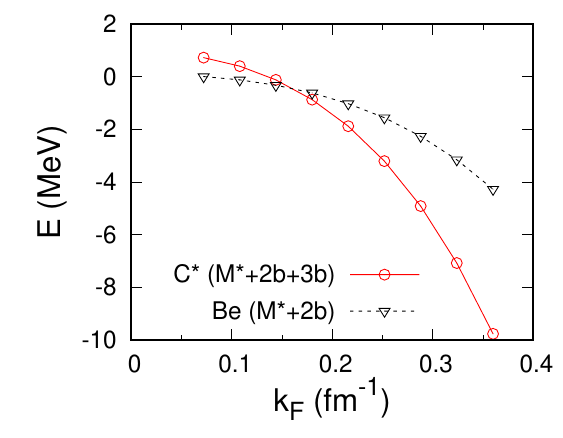}
\includegraphics[bb=0 0 162 126, width=0.5\linewidth]{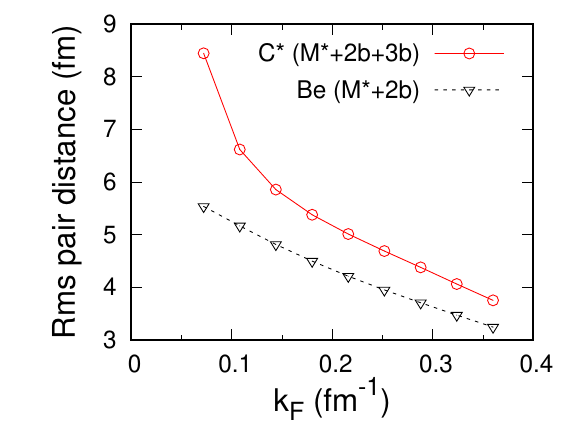}
\caption{Energies (left) and root-mean-square 
pair distance (right) of the ground state of $^8$Be
and the Hoyle state C$^*$ in neutron matter of the Fermi momentum $k_F$.
$M^*$, $2b$, and $3b$ denote calculations with the effective mass, induced 
two- and three-alpha interactions, respectively.}
\label{fig:EneDis}       % Give a unique label
\end{figure*}

The alpha cluster environment is characterized by the Fermi momentum 
of background neutrons $k_F$, which affects the effective mass,
induced two- and three-alpha interactions in the Hamiltonian.
Here we discuss the energies and structure of the ground state of $^8$Be and 
the Hoyle state as a function of $k_F$.
The left panel of Fig.~\ref{fig:EneDis} plots the ground state of $^8$Be and 
the Hoyle state energies as a function of $k_F$.
The energies of both the $^{8}$Be and the Hoyle state
become lower for larger $k_F$ due to
the short-range attraction of the induced two-alpha interaction.
Eventually, the ground state of $^8$Be becomes a bound state 
for $k_F \gtrsim 0.08~\mathrm{fm}^{-1}$ 
and the Hoyle state is also bound for $k_F \gtrsim 0.16~\mathrm{fm}^{-1}$,
which correspond to approximately $10^{-4}$ and $10^{-3}$
of the saturation density 0.16 fm$^{-3}$, respectively.
To see the structure of these alpha particle systems,
we plot in the right panel of Fig.~\ref{fig:EneDis} 
the root-mean-square (rms) pair distance of the ground state of $^8$Be and
the Hoyle state. 
Both the ground state of $^8$Be and the Hoyle state become smaller 
for larger $k_F$ monotonically.
This shrinkage occurs because the internal amplitude of the 
wave function is strongly modified by the induced interactions
mostly by the attractive induced two-alpha interaction.
It is interesting to note that the size of the Hoyle state 
in the cold neutron matter shrinks as
is consistent with the study of the
finite system, e.g., the $\alpha-\alpha-n$ 
cluster calculation~\cite{Lyu15}.

\section{Conclusion and prospects}

We have made precise calculations 
for few-alpha systems immersed in cold neutron matter.
The alpha particle is treated as a polaron, i.e.,
quasiparticle dressed by excitations of background neutrons.
We have found the possibility that
the two- and three-alpha systems become bound 
in cold dilute neutron matter.
The ground state of $^8$Be and the Hoyle state 
become bound
at $\approx$~$10^{-4}$ and $10^{-3}$ of the saturation density.
The binding of these fundamental light nuclear ingredients 
would have a significant effect on the modeling of 
stellar collapse and neutron star mergers, 
as well as relevant 
reaction rates for nucleosynthesis.
To realize these effects by the presence 
of the cold neutron matter in experiments, 
it is interesting to investigate $2\alpha + Xn$ 
and $3\alpha + Xn$ systems.
For this purpose, it is desirable to relate
alpha clusters interacting with a few neutrons
and with infinite neutron matter.
A detailed study along this direction is underway
and will be reported elsewhere soon.

\section{Acknowledgement}

This work was in part supported 
by JSPS KAKENHI Nos. 17K05445, 18K03635, 18H01211, 18H04569, 
18H05406, and 19H05140, and the Collaborative 
Research Program 2021, Information Initiative Center,
Hokkaido University.

%
% BibTeX or Biber users please use (the style is already called in the class, ensure that the "woc.bst" style is in your local directory)
% \bibliography{name or your bibliography database}

\end{document}